\documentclass[twocolumn,prl,showpacs,aps,superscriptaddress]{revtex4-2}

\usepackage{amsmath}
\usepackage{multirow}
\usepackage{graphicx} % for figures
\usepackage{color}
\usepackage[latin1]{inputenc}
\usepackage{enumitem}
\usepackage[french]{babel}
\begin{document}

\date{\today}

\title{Magnetic phase diagram of cuprates and universal scaling laws}

\author{Yves Noat}

\affiliation{Institut des Nanosciences de Paris, CNRS, UMR 7588 \\
Sorbonne Universit\'{e}, Facult\'{e} des Sciences et Ing\'{e}nierie, 4 place
Jussieu, 75005 Paris, France}

\author{Alain Mauger}

\affiliation{Institut de Min\'{e}ralogie, de Physique des Mat\'{e}riaux et
de Cosmochimie, CNRS, UMR 7590, \\ Sorbonne Universit\'{e}, Facult\'{e} des
Sciences et Ing\'{e}nierie, 4 place Jussieu, 75005 Paris, France}

\author{William Sacks$^*$}

\affiliation{Institut de Min\'{e}ralogie, de Physique des Mat\'{e}riaux et
de Cosmochimie, CNRS, UMR 7590, \\ Sorbonne Universit\'{e}, Facult\'{e} des
Sciences et Ing\'{e}nierie, 4 place Jussieu, 75005 Paris, France}

\affiliation{Research Institute for Interdisciplinary Science,
Okayama University, Okayama 700-8530, Japan}



\begin{abstract}

In this article we consider the magnetic field phase diagram of
hole-doped high-$T_c$ cuprates, which has been given much less
attention than the temperature diagram. In the framework of the {\it
pairon model}, we show that the two characteristic energies, the
pair binding energy (the gap $\Delta_p$) and the condensation energy
($\beta_c$) resulting from pair correlations, give rise to two major
magnetic fields, the upper critical field $B_{c2}$ and a second
field, $B_{pg}$, associated with the pseudogap (PG). The latter
implies a second length scale in addition to the coherence length,
characteristic of incoherent pairs. Universal scaling laws for both
$B_{c2}$ and $B_{pg}$ are derived: $B_{c2}$ scales with the critical
temperature, $B_{c2}/T_c\simeq 1.65$\,T/K, in agreement with many
experiments, and $B_{pg}$ has a similar scaling with respect to
$T^*$. Finally, Fermi arcs centered on the nodal directions are
predicted to appear as a function of
magnetic field, an effect testable experimentally.\\
\\ *Corresponding author\,: william.sacks@sorbonne-universite.fr
\end{abstract}

\maketitle

\subsection{Introduction}

Establishing the magnetic phase diagram as a function of doping is a
key step to grasp the fundamental mechanisms underlying
superconductivity in cuprates. Under an applied magnetic field, type
II superconductors satisfy general properties. Vortices appear above
the lower critical field, $B_{c1}$, and the typical size of the
vortex core is the superconducting (SC) coherence length $\xi_0$.
The latter is directly related to the upper critical field,
$B_{c2}$, through the relation $B_{c2}= \frac{\Phi_0}{2 \pi
\xi_0^2}$, where $\Phi_0 = h/(2e)$ is the magnetic flux quantum.

In cuprates the phase coherence length is in the nanometer range so
that the upper critical field is often too large to be measured, in
which case its estimation can be made using indirect extrapolation
methods. Such is the case, for instance, for
YBa$_2$Cu$_3$O$_{7-\delta}$ (YBCO) or
Bi$_2$Sr$_2$CaCu$_2$O$_{8+\delta}$ (BSCCO). This experimental
limitation might explain some of the discrepancies for $B_{c2}$
obtained in different experiments.

The upper critical field has been deduced from the specific heat by
different groups \cite{EPL_Wen2003,EPL_Wang2008} on
La$_{2-x}$Sr$_x$CuO$_4$ (LSCO) for different hole concentrations
($p$) from underdoped to overdoped sides of the critical $T_c$ dome.
The authors measured the dependence of the $\gamma$ coefficient as a
function of magnetic field, from which $B_{c2}$ can be obtained.
Their results indicate that $B_{c2}$ follows a dome as a function of
$p$, and hence follows the critical temperature. These findings are
in qualitative agreement with the high-field measurements of the
resistivity \cite{Natcom_Grissonnanche2014}. On the other hand,
there is an apparent contradiction with experiments by Kato et al.
\cite{JPSJ_Kato2024}, who estimate the upper critical field of
Bi$_{2+x}$Sr$_{2-x}$CaCu$_2$O$_{8+\delta}$ from the measured
irreversibility field. In this case, $B_{c2}$ qualitatively follows
a dome for small carrier concentrations, but then increases
monotonically to high values in the overdoped regime. Therefore, in
order to address these issues, it is important to clarify the key
parameters, i.e. the characteristic energies, lengths, and magnetic
fields of high-$T_c$ cuprates.

More than 30 years after the discovery of cuprates by Bednorz and
M\"{u}ller \cite{ZPhys_Bednorz1986}, key questions such as the pairing
`glue', the nature of the pseudogap, and the role of the gap
parameter, remain open issues. Above $T_c$, contrary to conventional
superconductors, the normal metallic state is not recovered.
Instead, the pseudogap (PG) state is found, characterized by a
lowering of the quasiparticle DOS, which persists up to the higher
temperature $T^*$ \cite{Nat_Hashimoto2014}.

\begin{figure}[t]
\includegraphics[trim=10 10 10 10, clip, width=9.0 cm]{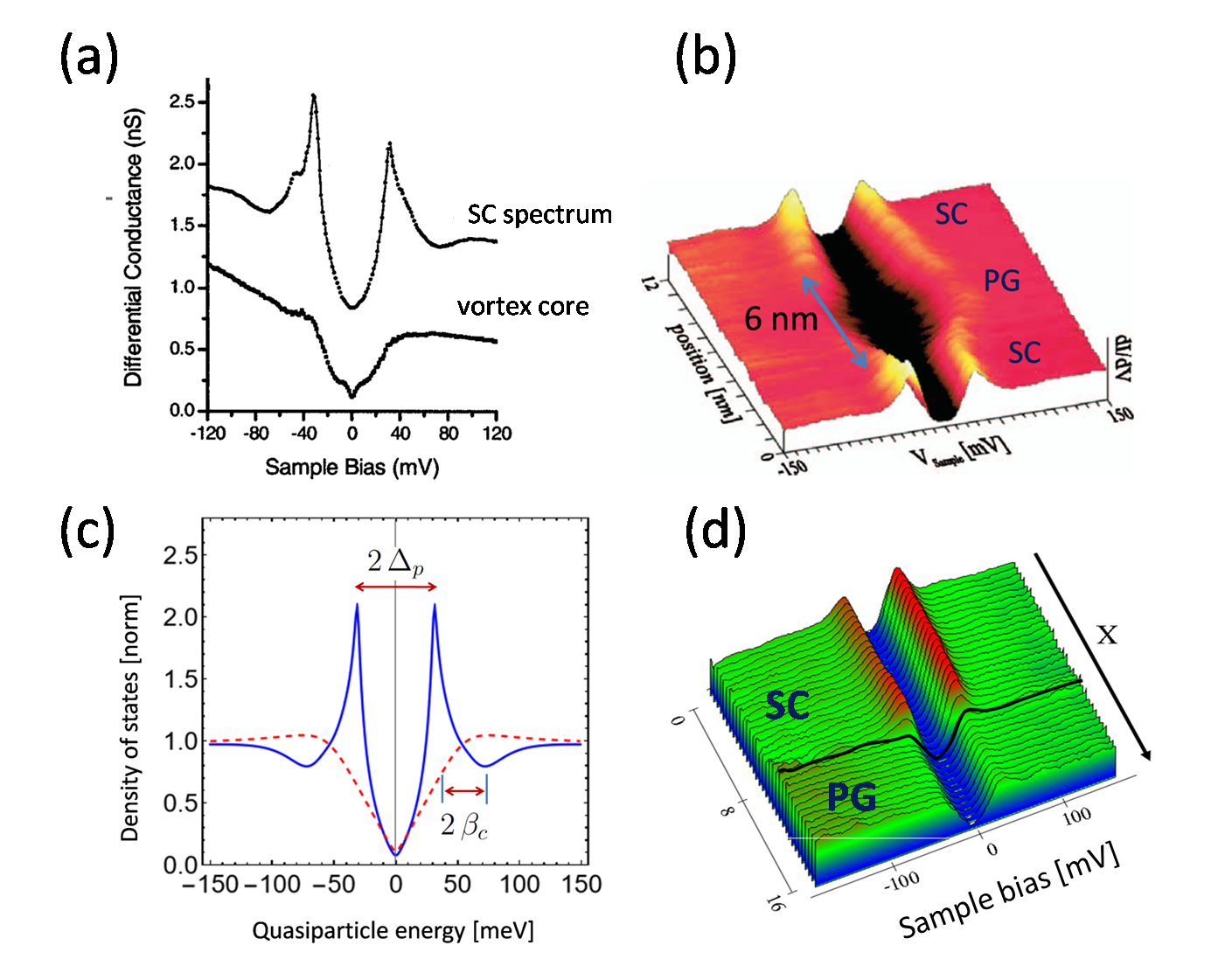}
\caption{(Color online) Strong experimental evidence of the
pseudogap (PG) state obtained when SC coherence is lost at {\it low
temperatures}. In contrast to the conventional case, this PG state
consists of incoherent pairons, even at the local
scale\cite{EPL_Sacks2017}. a) STS spectrum taken at low temperature
in the SC state as well as within the vortex core (adapted from
\cite{PRL_Pan2000}). b)  STS spectra taken along a line crossing a
vortex (adapted from \cite{Thesis_Kugler2000}). c) Theoretical
spectra (solid blue line) in the SC state and in the incoherent
pseudogap state, red dashed line (adapted from
\cite{ModelSimul_Noat2022, Jphys_Sacks2017, PRB_Sacks2006}. d) STS
spectra along a line crossing from an ordered region to a disordered
region (adapted from \cite{PRL_Cren2000}). All panels correspond to
BSCCO(2212).} \label{Fig_Vortex}
\end{figure}

In the BCS theory, the energy gap is proportional to the critical
temperature: $2\Delta/(k_B\,T_c) \simeq 3.52$. The situation is
different in cuprates since the amplitude of the gap is not directly
related to $T_c$. A gap of the same order of magnitude persists
above $T_c$ \cite{PRL_Renner1998} as well as within the vortex core,
see Fig.\,\ref{Fig_Vortex}, panel (a). This is confirmed by scanning
tunneling spectroscopy (STS) measurements in disordered BSCCO thin
films \cite{PRL_Cren2000}. In the SC state, the quasiparticle
spectrum displays the peak--dip structure. Crossing to the non-SC
region, the coherence peaks disappear while the gap is preserved
having roughly the same magnitude, see Fig.\,\ref{Fig_Vortex}, panel
(d). A similar transition is also observed as a function of
temperature \cite{PRL_Renner1998}. Therefore, since a gap exists
even when the superfluid density vanishes, it cannot be the order
parameter.

This conclusion is confirmed by the dependence of the energy gap
with doping. In fact, the gap, which is directly measured by
tunneling or photoemission spectroscopy, decreases linearly as a
function of carrier density
\cite{JPhysSocJap_Nakano1998,RepProgPhys_Hufner2008,Miyakawa PRL
1999}. In the underdoped regime, $T_c(p)$ has the opposite behavior
with respect to the energy gap since it {\it increases} with doping.
Two undecided questions thus emerge from these considerations:
\begin{enumerate}
\item  What is the order parameter in cuprates?
\item  What is the magnetic phase diagram and its connection to the temperature phase diagram?
\end{enumerate}
In this article, we establish important physical parameters of
hole-doped cuprates in the framework of the pairon model. Based on a
two--fluid Ginzburg-Landau approach, we obtain the expressions for
the characteristic lengths and magnetic fields. This allows to
deduce the magnetic phase diagram of cuprates, and to compare it
with the analogous temperature diagram.

% Figure 1
\subsection{Characteristic energy scales of cuprates}
% Figure 1
In order to answer the above questions, we consider the pairon model
wherein the SC state is the result of the condensation of preformed
pairs due to their mutual interaction \cite{SciTech_Sacks2015,
EPL_Sacks2017, PhysLettA_Noat2022}. Once SC coherence is lost due to
a magnetic field or rising temperature, the system becomes a
`glassy' PG state of incoherent pairons, such as in the vortex core.
The corresponding phase diagram for hole-doped cuprates can then be
described by two energy scales\,: the pairing energy gap $\Delta_p$
and the condensation (or correlation energy) $\beta_c$.

The meanings of $\Delta_p$  and $\beta_c$ in the pairon model are
distinct. Pairons are bound holes on adjacent copper sites which
form below $T^*$ as a result of their local antiferromagnetic
environment on the scale of $\xi_{AF}$, the AF correlation length
\cite{EPL_Sacks2017}.  They condense into a superconducting state at
the critical temperature $T_c$ ($T_c \leq T^*)$, forming a
collective pairon state. The energy (per pairon) in the pseudogap
state and in the SC state are respectively:
\begin{align}
E_{pg}  & =-\Delta_p\\
E_{SC}  &=-\Delta_p-\beta_c
\end{align}

It is important to note that the condensation energy $\beta_c$ is
due to correlations between pairons. In a sense it is a `hidden'
term since, contrary to conventional superconductors, it does not
correspond to a binding energy and is not measured as a usual gap in
photoemission or scanning tunneling spectroscopy. However the
coherence energy, $\beta_c$, is present in the quasiparticle
spectrum for energies just above the coherence peak characterized by
a sharp dip, see Fig.\,\ref{Fig_Vortex}(c). The precise analysis of
the quasiparticle spectra using a gap function $\Delta(E_k)$ reveals
quantitatively both energies $\Delta_p$ and $\beta_c$
\cite{SciTech_Sacks2015,PRB_Sacks2006,Jphys_Sacks2017}.

We have previously shown that both $\Delta_p$ and $\beta_c$ are
proportional to the same fundamental physical parameter, the
effective exchange energy $J_{eff}$ in the CuO plane. Note that
$J_{eff}$ is material dependent since it varies as a function of the
$c$-axis coupling between CuO planes. This can explain the
variations of $T_c$ between different materials.

As a result of topological constraints of holes in the CuO
plane\,\cite{PhysLettA_Noat2022}, the following relations are
derived for $\Delta_p$ and $\beta_c$ as a function of the hole
concentration:
\begin{align}
\Delta_p  &=J_{eff}(1-p')\\
\beta_c  &=J_{eff} p'(1-p')
\end{align}
Here $p'=(p-p_{min})/(p_{max}-p_{min})$ is the reduced density, with
$p_{min}=0.05$ and $p_{max}=0.27$. Thus, $p'$ conveniently ranges
from 0 to 1, from the start to end of the $T_c$ dome. Moreover, the
energy and temperature scales are linked by the relations
\cite{EPL_Sacks2017}:
\begin{align}
\Delta_p=2.2\,k_BT^* \\
\beta_c=2.2\,k_BT_c
%\nonumber
\label{Equa_Ener_Temp}
\end{align}
Considering the available data on BSCCO(2212), the numerical factor
does have some uncertainty:\,$\sim 2.25 \pm .05$. The simplest ratio
of the two energies therefore leads to\,:
\begin{align}
\frac{\beta_c}{\Delta_p}=\frac{T_c}{T^*}=p'
%\null &= p'
\end{align}
For LSCO this ratio is slightly lower ($\approx 0.8\,p'$) due to the
relatively smaller $T_c$. As shown in a previous article
\cite{Jphys_Sacks2018}, this ratio also determines the size of the
Fermi arcs at the critical temperature.

\subsection{Characteristic lengths}
% Figure 2
\begin{figure}
\hskip -9 mm\includegraphics[trim=10 10 10 10, clip, width=9.5
cm]{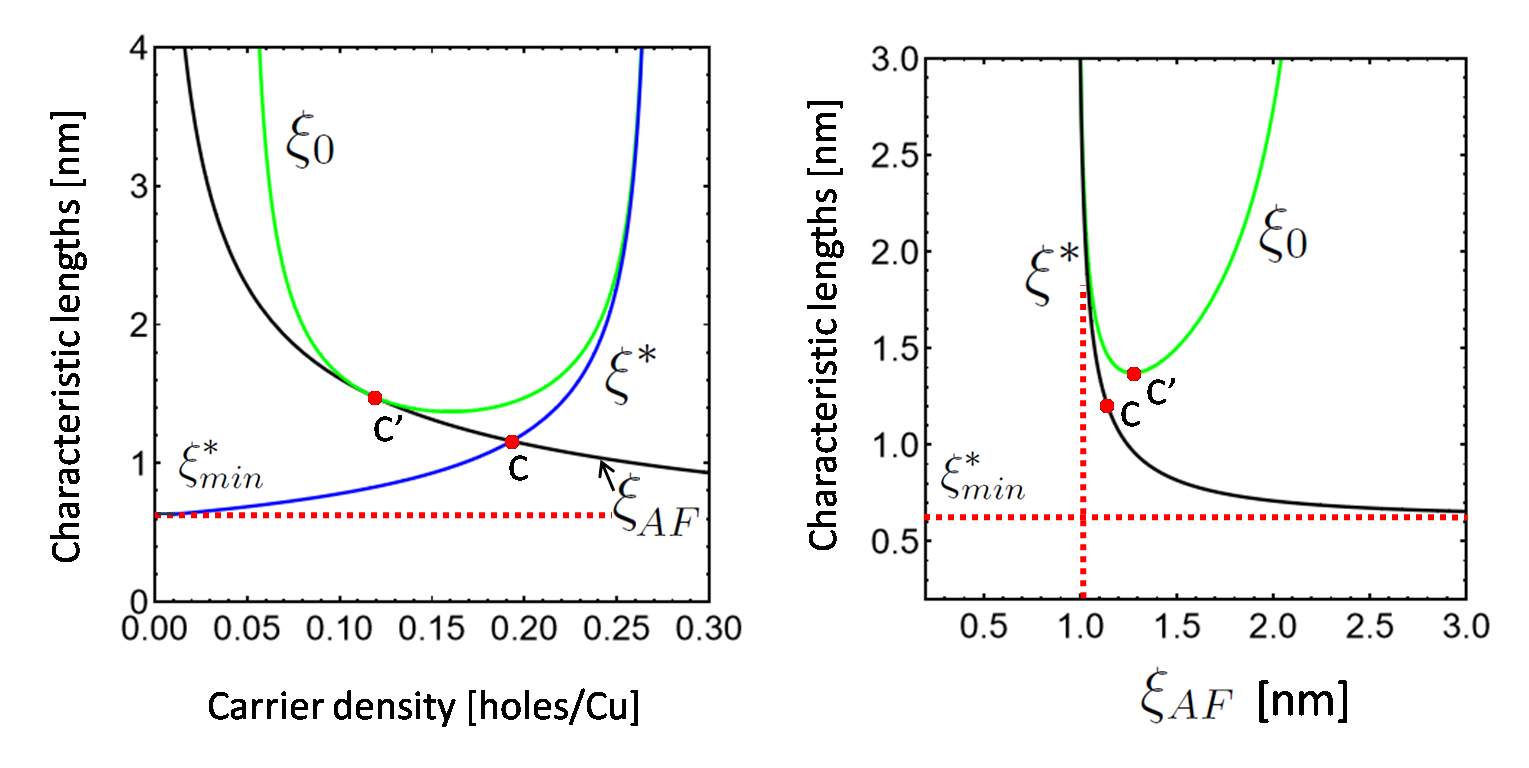} \caption{(Color online) Plot of the coherence
length $\xi_0$, the pair correlation length $\xi^*$, and the
antiferromagnetic correlation length $\xi_{AF}$, according to the
model.  a) All three lengths plotted as a function of carrier
concentration. b) $\xi_0$, $\xi^*$ plotted as a function of
$\xi_{AF}$. We note the special points c' and c, at the values
$p\simeq .12$, and $p \simeq .19$, respectively. The plots are
appropriate for BSCCO(2212).} \label{Fig_Lengths}
\end{figure}
A two-fluid model is a convenient way to derive the appropriate
length scales. In order to describe the two different fluids, the
superfluid (the coherent pairons of the condensate) and the fluid of
incoherent pairons (pairons excited out of the condensate), we
consider two coupled Ginzburg-Landau equations in zero applied
field\,:
\begin{align}
-\frac{\hbar^2}{2 m} \nabla^2 \psi_c + a \psi_c + b |\psi_c|^2 \psi_c &= 0 \quad \text{(condensate)} \\
-\frac{\hbar^2}{2 m} \nabla^2 \psi_{ex} + \bar{a} \psi_{ex} +
\bar{b} |\psi_{ex}|^2 \psi_{ex} &= 0 \quad \text{(excited pairs)}
\end{align}
Here $|\psi_c|^2 = n_s$ is the condensate density and $|\psi_c|^2 =
n_s$ is the excited pairs density. For example, qualitatively, the
condensate density $n_s$ gives rise to the typical STS spectrum
outside the vortex, whereas $n_{ex}$ aptly describes the local
density within the core (see Fig.\,\ref{Fig_Vortex} (b)).

Neglecting pair-breaking below $T_c$ and phase fluctuations above
$T_c$, the conditions that $n_s(T) + n_{ex}(T) = n_0$ for  $T \leq
T_c$ and  $n_s(T) =0$ for $T \geq T_c$ should be fulfilled.
%with $n_s(T_c)=n_0$ and $n_{ex}(T_c)=n_0$.
The variation of the free energy between the normal state and
superconducting state (assuming $T=0$) is
\begin{align}
\Delta F &= \frac{a^2}{2b} = \frac{1}{2} n_s |a| \\
&= \frac{1}{2} p' |a|
\end{align}
In addition, we have $|\psi_c|^2 = n_s(T) = |a|/b$. This allows to
compare to the expression of $\Delta F = \frac{1}{2} \beta_c p'$
expected in the pairon model, from which we deduce that $|a| =
\beta_c$. Including pair-breaking below $T_c$, relevant to the
highly-overdoped regime, gives the same result.

We therefore obtain the two length scales
$\xi_i=\sqrt{\frac{\hbar^2}{2m E_i}}$, with $i=1,2$, associated with
the two energies $E_1=\Delta_p$ and $E_2=|a|=\beta_c$:
\begin{align}
\xi^* &= \sqrt{\frac{\hbar^2}{2m \Delta_p}} \\
\xi_0 &= \sqrt{\frac{\hbar^2}{2m\beta_c}}
\end{align}
 Here $\xi^*$ is the length associated with the pairing gap
 while $\xi_0$ is the coherence length associated
 with superconducting phase coherence.

A third important length scale is the antiferromagnetic correlation
length and its dependence on the hole carriers, $\xi_{AF}(p)$. In a
simple approach \cite{{EPL_Sacks2017}}, we proposed that
$\xi_{AF}(p)$ is determined by the mean pairon-pairon distance in
the CuO plane\,:
\begin{align}
\xi_{AF}(p) &= a_0\sqrt{\frac{2}{p}}
\end{align}
where $a_0$ is the in-plane Cu-Cu distance. This formula is in
qualitative agreement with neutron experiments \cite{Birgeneau PRB
1988}. In this picture, a given pairon is surrounded by an area, a
Voronoi cell, corresponding to a well-defined AF magnetic order on
the scale of $\xi_{AF}(p)$. In \cite{{EPL_Sacks2017}}, we showed
that the pairon binding energy, $\Delta_p$, is proportional to the
Voronoi cell area, providing a simple explanation for its
$p$-dependence.

It is instructive to see that, after a little algebra, two of the
length scales $\xi^*$ and $\xi_{AF}(p)$ are connected by a simple
equation\,:
\begin{align}
\left(\frac{\xi^*_{min}}{\xi^*}\right)^2 +
\left(\frac{\xi^{min}_{AF}}{\xi_{AF}}\right)^2 = 1
\end{align}
where $\xi^*_{min}$ is the value of $\xi^*$ extrapolated to $p=0$,
and $\xi^{min}_{AF} = \xi_{AF}(.27)$. The above equation can be
represented by a hyperbola (solid black line
Fig.\,\ref{Fig_Lengths}, right panel). The third length, $\xi_0$,
can be conveniently expressed as\,:
$$
\xi_0 = \frac{1}{\sqrt{p'}}\,\xi^*
$$

The variation of the three length scales $\xi_0$, $\xi^*$ and
$\xi_{AF}$ as a function of carrier concentration are shown in
Fig.\,\ref{Fig_Lengths}, left panel. As seen in the figure, there is
a striking difference in the behavior of $\xi_0$ and $\xi^*$. In the
highly underdoped regime $\xi_0\gg \xi^*$, evidently the pairons
behave dominantly as bosons. However, in the overdoped regime, i.e.
towards the end of the $T_c$--dome, $\xi_0\rightarrow \xi^*$ as $p
\rightarrow p_{max}$, so the two length scales approach each other
asymptotically. At the same time, as can be seen using eqns.\,(3)
and (4), close to the critical point $p=p_{max}$ or $p'=1$, the
associated energy scales $\Delta_p$ and $\beta_c $, also approach
each other, and vanish identically in the limit $p \rightarrow
p_{max}$.

The overdoped side of the dome thus recalls the BCS case where the
coherence length is determined by the energy gap. However, it would
be misleading to conclude that one recovers the BCS mechanism in
this regime. Indeed, the underdoped and overdoped regimes clearly
correspond to different limits, given by the ratio $\xi_0/\xi^*$.
However, the mechanism at the heart of the cuprate SC coherence is
imposed by the conjunction of the geometrical constraints of the CuO
plane and the finite size of hole-pairs \cite{PhysLettA_Noat2022}.
Detailed considerations of the model confirm that pairons, i.e. two
holes bound on neighboring copper sites in an antiferromagnetic
environment, are the fundamental quantum objects across the
$T_c$--dome.

To conclude this discussion, we note that two special points in the
phase diagram exist\,: c' and c (respectively, at the concentrations
$p \simeq .12$ and $p \simeq .19$), see Fig.\,\ref{Fig_Lengths}.
These two points correspond to the common tangent of $\xi_{AF}$ with
$\xi_{0}$ and the intersection of $\xi_{AF}$ with $\xi^*$,
respectively. We see the further condition that $\xi_{AF}(p) \leq
\xi_{0}(p)$ for all $p$, and that c' (at $p\simeq .19$) is an
evident transition point involving the the characteristic length
$\xi^*$. The latter phenomenon will be tackled in a future report.

% Figure 3
\begin{figure}
\includegraphics[trim=10 10 10 10, clip, width=6.0 cm]{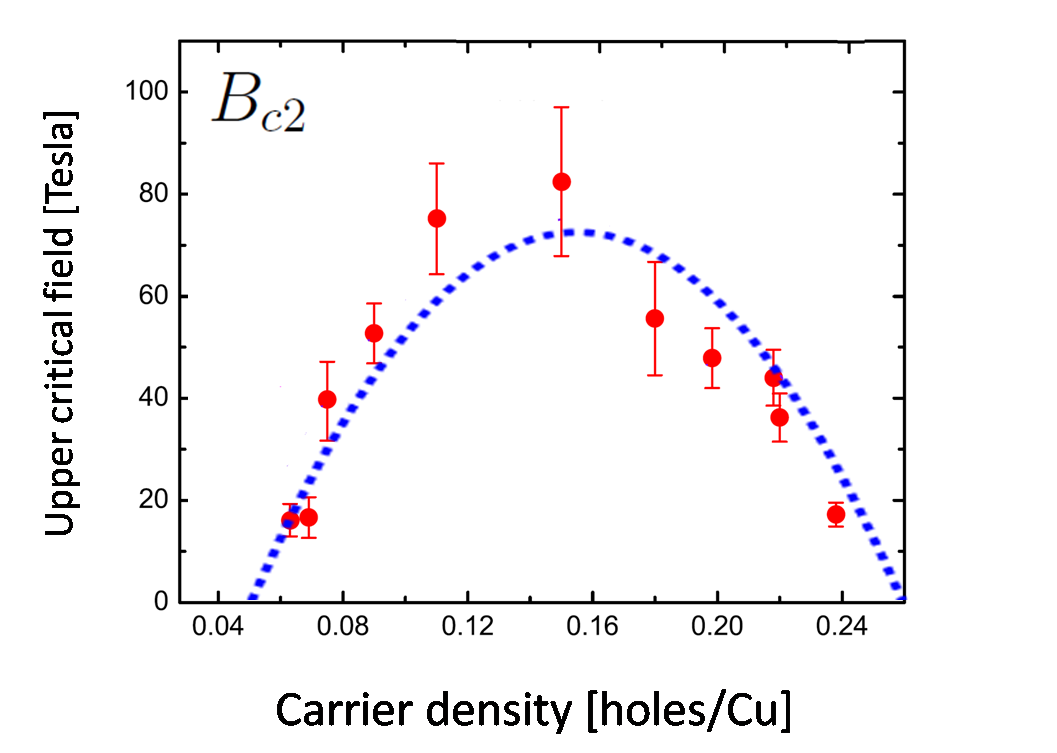} \caption{(Color
online) Upper critical field, $B_{c2}$, for LSCO as a function of
carrier density. Experimental values (red points) are taken from
Wang et  H.-H. Wen \cite{EPL_Wang2008}. Dashed blue line:
theoretical prediction from the model.} \label{Fig_Hc2_doping}
\end{figure}

\subsection{Characteristic magnetic fields of cuprates}

The coherence length $\xi_0 $, is the typical size of the vortex
core and it is associated with the upper critical field $B_{c2}$.
Similarly, $\xi^*$ is associated with a higher field $B_{pg}$
reflecting the pseudogap state. While $B_{c2}$ is the field
necessary to destroy SC phase coherence, we shall show that pairons
theoretically survive up to $B_{pg}>B_{c2}$.

From the two length scales, one can deduce the expressions of the
two important magnetic fields using the relations:
\begin{align}
B_{c2} &= \frac{\Phi_0}{2 \pi \xi_0^{2}} \\
B_{pg} &= \frac{\Phi_0}{2 \pi \xi^{*2}}
\end{align}
Replacing the expressions for $\xi_0$ and $\xi^*$, we obtain:
\begin{align}
B_{c2} & = \left(\frac{m \Phi_0}{\pi \hbar^2} \right) \beta_c \\
B_{pg} &= \left(\frac{m \Phi_0}{\pi \hbar^2} \right) \Delta_p
\end{align}
One first important consequence from Eq.\,(18) is that the upper
critical field must follow the superconducting dome. This is in good
agreement with the findings of Wang and Wen \cite{EPL_Wang2008}, as
shown in Fig.\,\ref{Fig_Hc2_doping}.

%\subsection{Universal scaling laws}
% Figure 4
\begin{figure}
\includegraphics[trim=10 10 10 10, clip, width=8. cm]{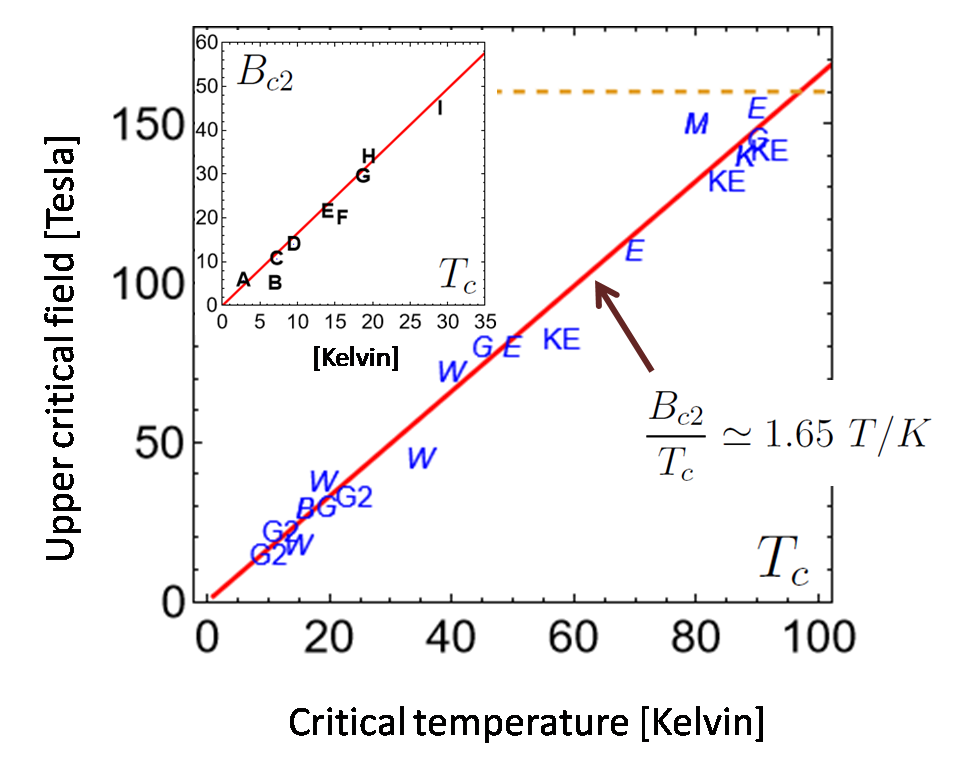}
\caption{(Color online) Upper critical field, $B_{c2}$, as a
function of the critical temperature, $T_c$, for different cuprate
materials and doping values. Inset\,: Similar plot for conventional
superconductors A to H, $B_{c2}$
follows a similar line with $B_{c2}/T_c\simeq 1.7$ T/K.\\
In the main plot\,:
 KE : Krusin Elbaum et al. BSCC0 \cite{PhysicaC_Krusin-Elbaum2003},
 K : Kugler et al. BSCCO \cite{Thesis_Kugler2000},
 M : Maggio Aprile et al. YBCO \cite{PRL_Maggio-Aprile1995},
 W : Wang et Wen LSCO \cite{EPL_Wang2008},
 G : Grisonnanche et al. YBCO \cite{Natcom_Grissonnanche2014},
 G2 : Grisonnanche et al. Tl2201 \cite{Natcom_Grissonnanche2014}.\\
Details for the inset:\\
 A : BaBi$_3$ \cite{SUST_Haldolaarachchige2014},
 B : NbSe$_2$ \cite{Spring_Nader2014},\cite{JLTPhys_Toyota1976},
 C : MgCNi$_3$ \cite{PhysicaC_Andrzejewski2007} ,
 D : NbTi \cite{IEEE_Lubell1983},
 E : V$_3$Ga \cite{PhysLett_Montgomery1966},
 F : V$_3$Si \cite{PRB_Orlando1979,CondMat_Khlopkin1999},
 G : Nb$_3$Sn \cite{PRB_Orlando1979},
 H : Nb$_3$Al \cite{PhysLettA_Foner1970},
 I : Rb$_3$C$_{60}$\cite{PRB_Shu1997}.}
\label{Fig_Hc2_vs_Tc}
\end{figure}
One can therefore deduce important scaling relations. Since the
upper critical field in the pairon model is proportional to the
coherence energy $\beta_c$, we have:
\begin{align}
\frac{B_{c2}}{\beta_c} &= \frac{m \Phi_0}{\pi \hbar^2} = \mathcal{C}
= \text{universal constant}
\end{align}
where $\mathcal{C}\simeq 8.6\,{\rm T/meV}$. Using $\beta_c =
2.25\,k_B\,T_c$, it follows that:
\begin{align}
\frac{B_{c2}}{T_c} \simeq 1.65\, {\rm T/K}
\end{align}
This relation should hold for any hole-doped cuprate and for
arbitrary carrier concentrations within the SC range. In
Fig.\,\ref{Fig_Hc2_vs_Tc}, we have plotted the upper critical field
for several materials and doping values. The agreement confirms the
linear relation between $B_{c2}$ and $T_c$. We note that a wide
variety of conventional superconductors follow a similar trend
(inset of Fig.\,\ref{Fig_Hc2_vs_Tc}), however in this case each
material has a fixed carrier concentration, in contrast to cuprates.
Finally, using Eqs. (5) and (19), the model predicts a similar
scaling of the pairing field with $T^*$\,: $B_{pg}/T^* \simeq 1.65$
T/K.

\subsection{Magnetic phase diagram and discussion}
% Figure 5
\begin{figure}
\includegraphics[trim=10 10 10 10, clip, width=7.5 cm]{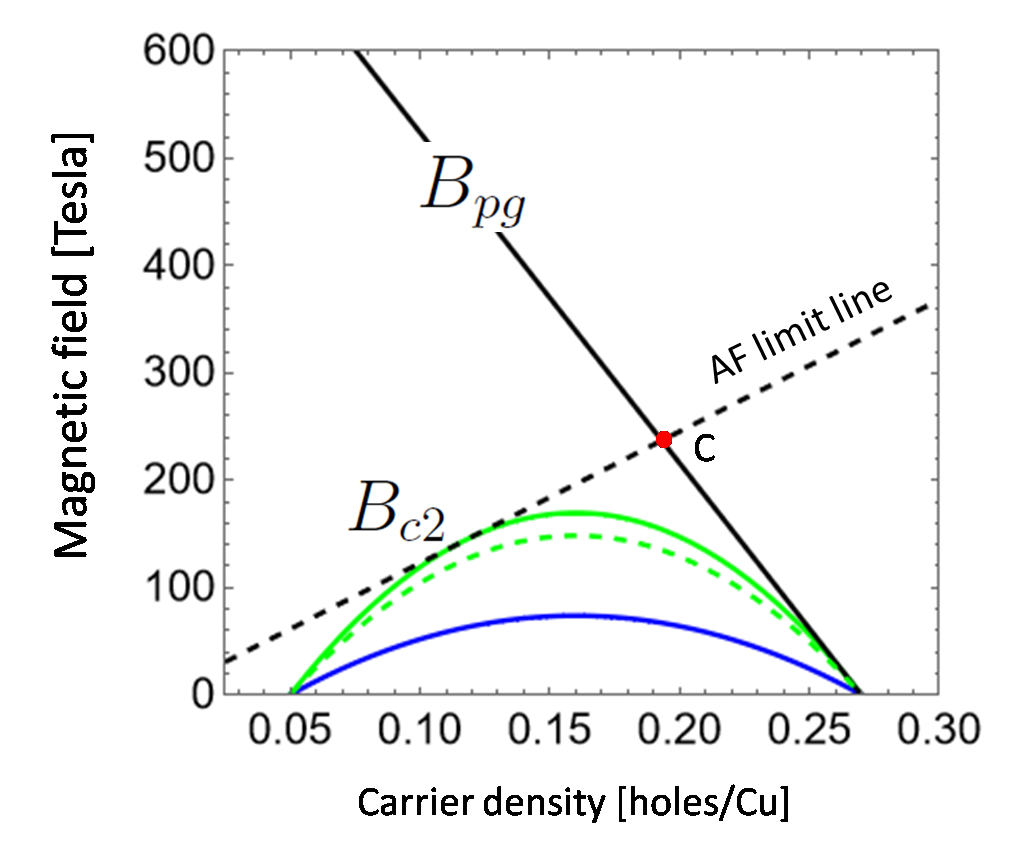}
\caption{(Color online) Magnetic phase diagram showing the two main
characteristic fields, the upper critical field $B_{c2}$ and the
field associated with pseudogap $B_{pg}$ (solid black line). Dashed
green line: $B_{c2}(p)$ deduced using the estimated value of the
coherence length from the vortex core in BSCCO
(Fig.\ref{Fig_Vortex}, panel (b)). Solid green line: $B_{c2}(p)$ of
BSCCO using the theoretical value for $\beta_c$. Solid blue line:
$B_{c2}(p)$ predicted for LSCO. Dashed black line: upper limit for
$B_{c2}(p)$ in the model (associated with the length scale
$\xi_{AF}$). Note that pairons form below the pseudogap magnetic
field, $B_{pg}$, and become coherent below $B_{c2}$.}
\label{Fig_Mag_Phasediag}
\end{figure}
The magnetic phase diagram thus appears to be very analogous to the
temperature phase diagram. Pairons form below the pseudogap field
$B_{pg}$ and become coherent below $B_{c2}$, see
Fig.\,\ref{Fig_Mag_Phasediag}. This is similar to the phase diagram
proposed by Krusin-Elbaum et al. \cite{PhysicaC_Krusin-Elbaum2003},
based on high magnetic field c-axis resistivity measurements. The
striking analogy between the temperature and magnetic phase diagrams
suggests that the magnetic field and the temperature produce similar
effects. Phase coherence is destroyed either by raising the
temperature above $T_c$, or by applying a magnetic field stronger
than $B_{c2}$, as in conventional superconductors. However, pairons
survive in cuprates above $T_c$ up to $T^*$ and, as our work
confirms, above $B_{c2}$ up to $B_{pg}$. The interpretation of
$B_{pg}$ as the pair-breaking field is thus reasonable, whereas
$B_{c2}$ breaks the phase coherence, in clear agreement with
experiment (Fig.\,\ref{Fig_Vortex}).

A unique property of cuprates is the existence of Fermi arcs of
`normal' electrons centered around the nodal directions at finite
$T$ (see \cite{Nat_Hashimoto2014} and Refs. therein). In our model
hole pairons are associated with a continuum of electron Cooper
pairs in momentum space. Four Fermi arcs then appear due to the
weaker binding energy of Cooper pairs with momenta close to the
nodal directions \cite{Jphys_Sacks2018}. As shown using ARPES
experiments \cite{Nat_Hashimoto2014}, a given Fermi arc is very
small at low temperature, in agreement with $d$-wave pairing. Then,
as $T \rightarrow T_c$, the Fermi arcs grow monotonically with
temperature, at a rate that is $p$-dependent. The complete Fermi
surface is recovered at the higher temperature $T^*$
\cite{Jphys_Sacks2018}.

To quantify the Fermi arc size using simple arguments, we recall
that the $d$-wave pair potential is well approximated by\,:
$\Delta(\theta)=\Delta_p\,\cos(2\theta)$ where $\theta$ is the
momentum angle measured from the antinodal direction. Thus, at some
finite temperature $T \leq T_c$, pairs whose energy satisfies
$\Delta_p\,\cos(2\theta) \leq 2.2\,k_B\,T$ will decay into normal
electron states around the node. The Fermi arc does indeed increase
with temperature, an effect which is more pronounced for higher
carrier concentrations. In particular, at $T=T_c$ we obtain the
critical angle $\theta_c$ which characterizes the width
(2$\theta_c$) of the Fermi arc\,:
\begin{align}
\cos(2\theta_c) = \frac{\beta_c}{\Delta_p}=\frac{T_c}{T^*}= p'
\end{align}
This critical angle has been directly measured by Hashimoto et al.
\cite{Nat_Hashimoto2014} where, for near optimal doping, the above
ratio is about 1/2 giving $\theta_c \simeq 30$°. (As previously
mentioned, the simple ratio above, valid for BSCCO-2212, would be
slightly smaller for LSCO by about 20\%.)

The same physical effect should also occur in a magnetic field.
Indeed, using eqns. (18) and (19) we obtain the analogous relation
for the critical angle\,:
\begin{align}
\cos(2\theta_c) = \frac{B_{c2}}{B_{pg}}
\end{align}
Again, the complete Fermi surface is only recovered at the higher
pairing field, $B_{pg}$. The existence of such Fermi arcs for the
lower field, $B \leq B_{c2}$, can be considered a critical test for
the pairon model.

\subsection{Conclusion}

In conclusion, we have derived expressions for the length scales and
magnetic fields of hole-doped cuprates. A two-fluid model, using two
coupled Ginzburg-Landau equations, describes the condensate (the
coherent pairs) and the incoherent pairs (pairs excited out of the
condensate). The two energy scales describing cuprates, the energy
gap, $\Delta_p$, and the condensation energy, $\beta_c$, give rise
to two characteristic magnetic fields, the upper critical field,
$B_{c2}$, and the pseudogap field, $B_{pg}$. The magnetic phase
diagram is therefore very analogous to the temperature phase
diagram. It appears that the same critical point, $p=p_{max}$, at
the end of the $T_c$ dome, also exists in the magnetic diagram.

In this work, we have established a universal scaling law for
$B_{c2}$ as a function of the critical temperature,
$B_{c2}/T_c\simeq 1.65$\,T/K, in quantitative agreement with
experiments, which should be valid for all cuprates on the
hole-doped side, and for any carrier concentration. A similar
scaling law is predicted for the pseudogap field, $B_{pg}$, but with
respect to $T^*$.

Instead of a single coherence length $\xi_0$, the model invokes a
second length, $\xi^*$, associated with the pairing gap. Their
dependence on the carrier concentration, from underdoped to
overdoped sides of the phase diagram, offers unique insights into
the cuprate mechanism. The present theory can be used to express the
thermodynamic field, the condensation energy, the lower critical
field $B_{c1}$, and the London penetration depth $\lambda$, which we
leave for a future report.

Finally, we propose that Fermi arcs should also exist as a function
of magnetic field, a prediction that can be used as a strong
experimental test of the model.

\subsection{Acknowledgements}
The authors gratefully acknowledge discussions with Dr. Hiroshi
Eisaki, Dr. Shigeyuki Ishida (AIST, Tsukuba), and Prof. Atsushi
Fujimori (Tokyo University).

A.M. and W.S. acknowledge partial support from the French National
Research Agency (ANR), project `Superstrong' under contract no.
ANR-22-CE30-0010 (Principal Investigator: Andrea Gauzzi).

W.S. is grateful for continual support of the RIIS Institute of
Okayama University, Japan, to Prof. Takayoshi Yokoya (host), and his
`visiting professor' status while on leave from SU.


\begin{thebibliography}{99}

% Bc2 measurements
\bibitem{EPL_Wang2008}
Y. Wang and H.-H. Wen, Doping dependence of the upper critical field
in La$_{2-x}$Sr$_x$CuO$_4$ from specific heat, Europhys. Lett., {\bf
81}, 57007 (2008).

\bibitem{EPL_Wen2003} H. H. Wen, H. P. Yang, S. L. Li, X. H. Zeng, A. A. Soukiassian, W. D. Si and X. X. Xi,
Hole doping dependence of the coherence length in La$_{2-x}$Sr$_x$CuO$_4$ thin films,  Europhys. Lett., {\bf  64 }, 790 (2003).

\bibitem{Natcom_Grissonnanche2014} G. Grissonnanche, O. Cyr-Choini\`ere, F. Lalibert\'e, S. Ren\'e de Cotret, A. Juneau-Fecteau, S. Dufour-Beaus\'ejour, M. -\`E. Delage, D. LeBoeuf, J. Chang, B. J. Ramshaw, D. A. Bonn, W. N. Hardy, R. Liang, S. Adachi, N. E. Hussey, B. Vignolle, C. Proust, M. Sutherland, S. Kr\"{a}mer, J. -H. Park, D. Graf, N. Doiron-Leyraud and Louis Taillefer, Direct measurement of the upper critical field in cuprate superconductors, Nature Communications {\bf 5}, Article number: 3280 (2014).

\bibitem{JPSJ_Kato2024} Junichiro Kato, Shigeyuki Ishida, Tatsunori Okada, Shungo Nakagawa, Yutaro Mino,
Yoichi Higashi , Takanari Kashiwagi , Satoshi Awaji , Akira Iyo ,
Hiraku Ogino, Yasunori Mawatari, Nao Takeshita, Yoshiyuki Yoshida,
Hiroshi Eisaki, and Taichiro Nishio, Doping dependence of upper
critical field of high-$T_c$ cuprate
Bi$_{2+x}$Sr$_{2-x}$CaCu$_2$O$_{8+\delta}$ estimated from
irreversibility field at zero temperature, J. Phys. Soc. Jpn. {\bf
93}, 104705 (2024).

% Intro

\bibitem{ZPhys_Bednorz1986} J. G. Bednorz, K. A. M\"{u}ller, Possible high $T_c$ superconductivity in the Ba--La--Cu--O system , Zeitschrift f\"{u}r Physik B Condensed Matter {\bf 64}, 189 (1986).

% pseudogap

\bibitem{Nat_Hashimoto2014} Makoto Hashimoto, Inna M. Vishik, Rui-Hua He, Thomas P. Devereaux and Zhi-Xun Shen, Energy gaps in high-transition-temperature cuprate superconductors, Nature Physics {\bf 10}, 483 (2014).

\bibitem{PRL_Renner1998} Ch. Renner, B. Revaz, J.-Y. Genoud, K. Kadowaki, and
 \o. Fischer, Pseudogap precursor of the superconducting gap in under- and overdoped
Bi$_2$Sr$_2$CaCu$_2$O$_{8+\delta}$, Phys. Rev. Lett. {\bf 80}, 149 (1998).

\bibitem{PRL_Cren2000} T. Cren, D. Roditchev, W. Sacks, J. Klein, J.-B. Moussy, C. Deville-Cavellin,
and M. Lagu\"{e}s, Influence of disorder on the local density of states
in high-$T_c$ superconducting thin films, Physical Review Letters
{\bf 84}, 147 (2000).

\bibitem{JPhysSocJap_Nakano1998} Tohru Nakano, Naoki Momono, Migaku Oda, and Masayuki Ido,
Correlation between the doping dependences of superconducting gap
magnitude $2\Delta_0$ and pseudogap temperature $T^*$ in high-T$_c$
cuprates , J. Phys. Soc. Jpn. {\bf 67}, 2622-2625 (1998).

\bibitem{RepProgPhys_Hufner2008} S. H\"ufner, M. A. Hossain, A Damascelli, and G. A. Sawatzky,Two gaps make a high-temperature superconductor?,
Rep. Prog. Phys., {\bf 71}, 062501 (2008).

\bibitem{Miyakawa PRL 1999}
N. Miyakawa, J. F. Zasadzinski, L. Ozyuzer, P. Guptasarma, D. G.
Hinks, C. Kendziora, and K. E. Gray, Predominantly superconducting
origin of large energy gaps in underdoped
Bi$_2$Sr$_2$CaCu$_2$O$_{8+\delta}$ from tunneling spectroscopy,
Phys. Rev. Lett. \textbf{83}, 1018 (1999).

% Figures Spectres et vortex

\bibitem{PRL_Pan2000}S. H. Pan, E. W. Hudson, A. K. Gupta, K.-W. Ng, H. Eisaki, S. Uchida, and J. C. Davis, STM Studies of the Electronic Structure of Vortex Cores in Bi$_2$Sr$_2$CaCu$_2$O$_{8+\delta}$, Phys. Rev. Lett.  {\bf 85}, 1536 (2000).

\bibitem{Thesis_Kugler2000} Kugler, M., 2000, Ph.D. thesis, University of Geneva.

\bibitem{ModelSimul_Noat2022} Yves Noat  Alain Mauger, William Sacks,
Statistics of the cuprate pairon states on a square lattice,
Modelling Simul. Mater. Sci. Eng. {\bf 31}, 075010 (2023).

%\bibitem{EPL_Cren2000} T. Cren, D. Roditchev, W. Sacks and J. Klein,
%Constraints on the quasiparticle density of states in high-Tc superconductors, , Europhys. Lett. {\bf 52}, 203 (2000).

\bibitem{Jphys_Sacks2017} William Sacks, Alain Mauger and Yves Noat, Universal spectral signatures in pnictides and cuprates: the role of quasiparticle-pair coupling, J. Phys.: Condens. Matter  {\bf 29}, 445601 (2017).

\bibitem{PRB_Sacks2006} W. Sacks, T. Cren, D. Roditchev, and B. Dou\c{c}ot, Quasiparticle spectrum of the cuprate Bi$_2$Sr$_2$CaCu$_2$O$_{8+\delta}$: Possible connection to the phase diagram, Phys. Rev. B  {\bf 74}, 174517(2006).

% Two energy scales in the pairon model

\bibitem{SciTech_Sacks2015}  W. Sacks, A. Mauger, Y. Noat, Pair\,--\,pair interactions as a mechanism for
high-T$_c$ superconductivity, Superconduct. Sci. Technol., {\bf 28}
105014 (2015).

\bibitem{EPL_Sacks2017} W. Sacks, A. Mauger and Y. Noat, Cooper pairs without glue in high-$T_c$ superconductors: A universal phase diagram, Euro. Phys. Lett {\bf 119}, 17001 (2017).

\bibitem{PhysLettA_Noat2022} Yves Noat, Alain Mauger, William Sacks. Superconductivity in cuprates governed by topological constraints. Physics Letters A {\bf 444}, 128227 (2022).

\bibitem{Jphys_Sacks2018} William Sacks, A. Mauger and Y. Noat, Origin of the Fermi arcs in cuprates: a dual role of quasiparticle and pair excitations, Journal of Physics: Condensed Matter, {\bf 30},  475703 (2018).

\bibitem{Birgeneau PRB 1988}
R. J. Birgeneau, D. R. Gabbe, H. P. Jenssen, M. A. Kastner, P. J.
Picone, T. R. Thurston, G. Shirane, Y. Endoh, M. Sato, K. Yamada, Y.
Hidaka, M. Oda, Y. Enomoto, M. Suzuki, and T. Murakami,
Antiferromagnetic spin correlations in insulating, metallic, and
superconducting La$_{2-x}$Sr$_x$CuO$_4$ Phys. Rev. B \textbf{38},
6614 (1988).

% Bc2 cuprates

\bibitem{PhysicaC_Krusin-Elbaum2003} L. Krusin-Elbaum, T. Shibauchi, G. Blatter, C.H. Mielke, M. Li, M.P. Maley, P.H.Kes,
Pseudogap state in overdoped Bi$_2$Sr$_2$CaCu$_2$O$_{8+y}$, Physica C {\bf 387}, Pages 169-174 (2003).

\bibitem{PRL_Maggio-Aprile1995} Maggio-Aprile, I., C. Renner, A. Erb, E. Walker, and {\O}. Fischer,
Direct vortex lattice imaging and tunneling spectroscopy of flux
lines on YBa$_2$Cu$_3$O$_{7-\delta}$, Phys. Rev. Lett. {\bf 75},
2754 (1995).

% Bc2 conventionnal SC

\bibitem{SUST_Haldolaarachchige2014} Neel Haldolaarachchige, S. K. Kushwaha, Quinn Gibson and R. J. Cava,
Superconducting properties of BaBi$_3$, Supercond. Sci. Technol.
{\bf 27} 105001 (2014).

\bibitem{Spring_Nader2014} Adel Nader and Pierre Monceau, Critical field of 2H-NbSe$_2$ down to 50mK, SpringerPlus, {\bf 3}, 16 (2014).

\bibitem{JLTPhys_Toyota1976} N. Toyota, H. Nakatsuji, K. Noto, A. Hoshi, N. Kobayashi, and Y. Muto
and Y. Onodera, Temperature and angular dependences of upper
critical fields for the layer structure superconductor 2H-NbSe$_2$,
Journal of Low Temperature Physics, {\bf 25}, Nos. 3/4, 485 (1976).

\bibitem{PhysicaC_Andrzejewski2007} Bart{\l}omiej Andrzejewski, Tomasz Klimczuk, Robert J. Cava, The upper critical field in doped MgCNi$_3$,
Physica C, Volumes {\bf 460-462}, 706 (2007).

\bibitem{IEEE_Lubell1983} M. Lubell, Empirical scaling formulas for critical current and critical field for commercial NbTi,
 IEEE Transactions on Magnetics {\bf  19}, 754  (1983).


\bibitem{PhysLett_Montgomery1966} D.B. Montgomery, H. Wizgall, Measurement of the upper critical field vanadium - gallium alloys,
Physics Letters {\bf 22}, 48 (1966).

\bibitem{PRB_Orlando1979} T. P. Orlando, E. J. McNiff, Jr., S. Foner, and M. R. Beasley, Critical fields, Pauli paramagnetic limiting, and material parameters of Nb$_3$Sn and V$_3$Si, Phys. Rev. B {\bf 19}, 4545 (1979).

\bibitem{CondMat_Khlopkin1999}
M. N. Khlopkin, Anisotropy of the upper critical field and specific heat in V$_3$Si - a superconductor with cubic symmetry,
Condensed Matter  {\bf 69}, 26 (1999).

\bibitem{PhysLettA_Foner1970} S. Foner, E.J. McNiff Jr., B.T. Matthias, T.H. Geballe, R.H. Willens, E. Corenzwit, Upper critical fields of high-temperature superconducting Nb$_{1-y}$(Al$_{1-x}$Ge$_x$)$_y$ and Nb$_3$Al: Measurements of H$_{c2} > 400$ kG at 4.2$^\circ$K
Physics letters. A, {\bf  31}, 349 (1970).

\bibitem{PRB_Shu1997} Shaoyan Chu and Michael E. McHenry, Synthesis and superconducting properties of a Rb$_3$C$_{60}$ single crystal, Phys. Rev. B  {\bf 55}, 11722 (1997).
%\bibitem{PhysicaC_Hou1994} J.G. Hou, X.-D. Xiang, Vincent H. Crespi, Marvin L. Cohen, A. Zettl,
%Magnetotransport in single-crystal Rb$_3$C$_6o$, Physica C  {\bf 228}, 175  (1994).

% Biblio Magnetic phase diagram (short)

\end{thebibliography}
\end{document}